\documentclass[12pt]{article}
\usepackage{multicol}
\title{Multicols Demo}
\usepackage{graphicx} 
\usepackage[toc,title,page]{appendix}
\usepackage{cite} 
\usepackage{amsfonts}
\usepackage{amsmath}
\usepackage{amssymb}
\usepackage{physics}
\usepackage[dvipsnames]{xcolor}
\usepackage{feynmp-auto}
\usepackage{epsfig}
\usepackage{slashed}
\allowdisplaybreaks
\usepackage{comment}
\usepackage{authblk}
\usepackage{blindtext}
\usepackage{hyperref}

\newcommand{\bc}{\begin{center}}
\newcommand{\ec}{\end{center}}
\newcommand{\me}{\medskip}
\renewcommand{\bf}[1]{\mathbf{#1}}

\title{\textbf{Entanglement Entropy of Compton Scattering with a Witness
}}
\author{Shanmuka Shivashankara\footnote{sshivashankara@colgate.edu}}
\affil{\itshape\small Colgate University, Department of Physics and Astronomy, \\ \itshape\small Hamilton, NY  USA}
\date{May 2023}

\numberwithin{equation}{section}

\begin{document}
\maketitle


\begin{abstract}
Unitarity and the optical theorem are used to derive the reduced density matrices of Compton scattering in the presence of a witness particle.  Two photons are initially entangled wherein one photon participates in Compton scattering while the other is a witness, i.e. does not interact with the electron.  Unitarity is shown to require that the entanglement entropy of the witness photon does not change after its entangled partner undergoes scattering. The final mutual information of the electronic and witness particle's polarization is nonzero for low energy Compton scattering.  This indicates that the two particles become correlated in spite of no direct interaction. Assuming an initial maximally entangled state, the change in entanglement entropy of the scattered photon's polarization is calculated in terms of Stokes parameters.  A common ratio of areas occurs in the final reduced density matrix elements, von Neumann entropies, Stokes parameter, and mutual information.  This common ratio consists of the Thomson scattering cross-section and an accessible regularized scattering area. 
\end{abstract}

\section{Introduction}

Recent literature has produced quantum field theoretic calculations of representative scalar and electromagnetic interactions in the area of quantum information science (QIS) \cite{Seki} - \cite{Araujo2}.  The algorithm for calculating the QIS metrics starts with specifying an inital entangled state in terms of degrees of freedom such as the particles' polarizations and momenta.  Then, the inital and final reduced density matrices can be calculated. 
From the latter, the von Neumann entanglement entropy is used to calculate the mutual information of degrees of freedom.  The entropy and mutual information measure the degree of maximal knowledge and correlation between degrees of freedom, respectively.  A recent review of QIS for particle physicists is provided in \cite{Lykken}.\me  

Calculations of the above metrics also occur for a witness particle in \cite{Araujo1,Araujo2}.  A witness particle is initially entangled with one of the two scattering particles but does not participate in the interaction.  It was found that the reduced density matrix of the witness particle changes due to the scattering.  This means the witness particle's information is affected by the scattering although it did not partake directly in the interaction.\me  

Herein, an electromagnetic interaction is considered. A photon is assumed to be entangled with another photon that subsequently partakes in Compton scattering.  The following work is distinguished from the above work by keeping unitarity, i.e. not dropping the forward scattering amplitudes \cite{Seki}-\cite{Araujo2}.  This inclusion gives rise to different results for the QIS metrics such as the entropy and mutual information.\me

In section \ref{witness}, unitarity is shown to leave the witness particle's reduced density matrix and von Neumann entropy unchanged by the scattering event.  This means unitarity does not allow the witness particle to gain or lose information via the scattering event between the electron and the other photon.  This holds for an initially pure or mixed state.  This contrasts with past work, wherein unitarity is not maintained, and the witness particle's reduced density matrix changes\cite{Araujo1,Araujo2}.\me

In section \ref{gamma and x}, the reduced density matrix of the scattered photon's polarization in terms of Feynman amplitudes is derived.  This calculation is an alternative to the literature in that unitarity is upheld and the optical theorem is utilized.  Unitarity leaves the density matrix's normalization unchanged after the interaction. The algorithm presented is readily adapted to other scattering processes.  From the density  matrix, the QIS metrics can be calculated.  The change in the von Neumann entanglement entropy of the scattered photon is evaluated in terms of a Stokes parameter in section \ref{gamma} assuming an initial maximally entangled mixed state.  The final mutual information of the electronic and witness particle's polarization is nonzero. This means that the electron and witness particle become entangled in spite of not interacting directly.  The entanglement entropy, Stokes parameter, mutual information and all the coefficients of the reduced density matrix elements are seen to have a common ratio of areas.\me

\section{No change in witness particle's entropy}\label{witness}

\begin{flushleft}

\begin{figure}[ht] \label{feynman}
\bc
\includegraphics{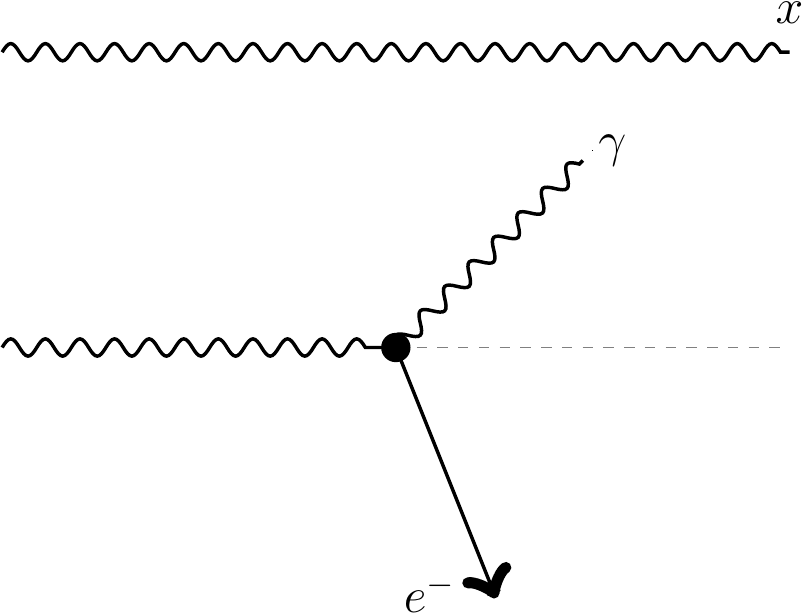}
\caption{$\gamma$ scatters off an $e^-$ at rest. $x$ is a witness particle that does not participate in the interaction.}
    \ec
\end{figure}

The initial pure state \cite{Fan1} is\medskip

\bc
$|i\rangle = |e^-,r\rangle$ (cos $\eta |\gamma, \uparrow \rangle |x,\downarrow \rangle +e^{i\beta}$ sin $\eta\ |\gamma,\downarrow\rangle |x,\uparrow\rangle),$
\ec \medskip

where $\eta$ is the entanglement parameter for particles $\gamma$ and $x$.  $e^{i\beta}$ is a relative phase.  $|e^-,r\rangle$ is an electron state with spin $r$, $|\gamma, \uparrow\rangle$ is a spin up photon, and $|x,\downarrow\rangle$ is a spin down photon. The momenta and a normalization constant are suppressed. The $e^-$ and $\gamma$ scatter while $x$ does not interact (see Figure \ref{feynman}).  $x$ is a spectator particle or witness to the interaction. The following shows that the reduced density matrix of the witness particle before and after the scattering is the same.  This implies there will be no change in particle $x$'s von Neumann entanglement entropy, $S_{EE} = -tr(\rho \log\rho)$, due to the scattering.\medskip

The initial density matrix is \medskip

\bc
$\rho^i = |i\rangle\langle i|$.\medskip
\ec

The final density matrix after the scattering event is\medskip

\bc 
$\rho^f = S \rho^i S^\dag = S|i\rangle\langle i|S^\dag$
\ec \medskip
where $S$ is unitary or $SS^\dag = 1$. \medskip

Using the completeness relation\medskip 
\[I= \sum_{s_1,s_2} |e^-,s_1;\gamma,s_2\rangle\langle e^-,s_1;\gamma,s_2|,\] \medskip

the reduced density matrix of particle $x$ is\\
\end{flushleft}

\begin{align*}
\rho^f_{x}&=tr_{e-,\gamma}(I\rho^f) = tr_{e-,\gamma}(\sum_{s_1,s_2} |e^-,s_1;\gamma,s_2\rangle\langle e^-,s_1;\gamma,s_2| \rho^f)\\
 =&\sum \langle e^-,s_1;\gamma,s_2|\ p_f\ |e^-,s_1;\gamma,s_2\rangle\\
  &\\
 =&\sum \langle e^-,s_1;\gamma,s_2|\ S|i\rangle\langle i|S^\dagger\ |e^-,s_1;\gamma,s_2\rangle\\
 &\\
 =&\ \ \ \cos^2 \eta |x,\downarrow \rangle\langle x,\downarrow |  \sum\langle e^-,s_1;\gamma,s_2|S|e^-,r;\gamma,\uparrow \rangle  \langle e^-,r;\gamma,\uparrow|S^\dagger |e^-,s_1;\gamma,s_2 \rangle \\
 &+\sin^2 \eta |x,\uparrow \rangle\langle x,\uparrow |\sum\langle e^-,s_1;\gamma,s_2|S|e^-,r;\gamma,\downarrow \rangle  \langle e^-,r;\gamma,\downarrow|S^\dagger |e^-,s_1;\gamma,s_2 \rangle \\
&+ \cos \eta\ \sin \eta\  e^{i\beta}|x,\uparrow\rangle\langle x,\downarrow|*\\
&\indent \sum\langle e^-,s_1;\gamma,s_2|S|e^-,r;\gamma,\downarrow \rangle  \langle e^-,r;\gamma,\uparrow|S^\dagger |e^-,s_1;\gamma,s_2 \rangle \\
&+ \cos \eta\ \sin \eta\ e^{-i\beta}|x,\downarrow\rangle\langle x,\uparrow|*\\
&\indent \sum\langle e^-,s_1;\gamma,s_2|S|e^-,r;\gamma,\uparrow \rangle  \langle e^-,r;\gamma,\downarrow |S^\dagger |e^-,s_1;\gamma,s_2 \rangle
\end{align*}

Notice each term in the latter expression can be simplified.  By reversing the order of the amplitudes, using the completeness relation, and $S^\dagger S = 1$, only the  $\cos^2\eta$ and $\sin^2\eta$  terms are nonzero.  Finally,\smallskip
\begin{align}\label{reducedx}
  \rho^f_{x} &= \cos^2\eta|x,\downarrow \rangle \langle x, \downarrow | + \sin^2\eta |x,\uparrow \rangle \langle x, \uparrow |\\
 &= \rho^i_{x}\nonumber\end{align}
More generally, if the completeness relation includes three particle states (double-Compton scattering and/or pair production) in addition to two particle states, equation \ref{reducedx} would still hold.  Since the reduced density matrix of the witness particle $x$ is the same both before and after the scattering, there is no change in particle $x$'s von Neumann entanglement entropy, 
\begin{align*}
\Delta S_{EE} = -tr(\rho^f_x \log\rho^f_x) + tr(\rho^i_x \log\rho^i_x)=0.
\end{align*}
 This would not be true without unitarity \cite{Araujo1}.  If $\rho^i$ is a mixed state instead of a pure state, e.g. by dropping the off-diagonal (coherence) terms $\cos\eta \sin\eta\ e^{\pm i\beta}$, the reduced density matrix of $x$ still remains unchanged by the scattering.  The electron's polarization and the type of interaction, Compton or otherwise, also does not change the result.    

\section{Initial photonic entanglement  entropy and mutual information}

This section covers the initial quantum information metrics, which are needed for section \ref{gamma} where the change in entropy and mutual information are calculated.  Let an initial reduced density matrix for the photonic polarizations be
\begin{align}\label{imixed}
\rho^i_{\gamma, x} &=\  \cos^2\eta |\uparrow,\downarrow \rangle \langle \uparrow, \downarrow | + 
\sin^2\eta |\downarrow,\uparrow \rangle \langle \downarrow, \uparrow |
\end{align}\me
The latter is a mixed state, i.e. $tr\rho^2 \neq 1.$  The expected inital spin of $\gamma$ along the collision axis \footnote{$\sigma_z = | \uparrow \rangle \langle \uparrow | - | \downarrow \rangle \langle \downarrow|$ where $\uparrow$ and $\downarrow$ refer to right and left-handed circularly polarized light, respectively.} and the initial von Neumann entanglement entropy between $x$ and $\gamma$ are
\begin{align*}
\langle\sigma_z \rangle &= tr(\sigma_z\ \rho^i_\gamma) = \cos\ 2\eta,\\
&\\
S^i_{EE} &= -tr(\rho^i_{\gamma,x}\ \log\ \rho^i_{\gamma,x})\\
&= -(\cos^2\eta\ \log\cos^2\eta\ +\ \sin^2\eta\ \log\sin^2\eta)\\
&=S^i_{EE,\gamma}=S^i_{EE,x},\end{align*}
where $S^i_{EE,\gamma}$ and $S^i_{EE,x}$ are the initial von Neumann entanglement entropies of particles $\gamma$ and $x$, respectively. For a pure state, $S^i_{EE}$ would be zero, representing maximal knowledge of the system of two polarizations.\\

\begin{figure}[ht]
\begin{center}
\includegraphics[width=\textwidth]{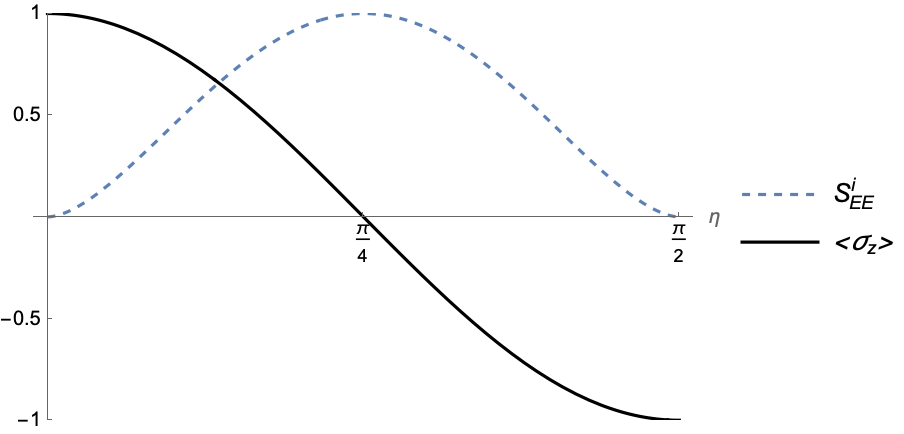}
    \caption{$S^i_{EE}$ is the initial entanglement entropy of $\gamma$ and $x$'s polarization.\\
    $\langle \sigma_z \rangle$ is the initial expected spin of $\gamma$ along the collision axis.  The software Mathematica plots $S^i_{EE}$ correctly after writing log base 2 in terms of log base $e$.}
    \label{see}
\end{center}
\end{figure}

Notice log has base 2 in reference to the two possible polarizations of each of the two bits ($\gamma$ and $x$).  In Figure \ref{see} at $\eta = \pi/4$, $S^i_{EE} = \log 2 = 1$ bit, representing not knowing which of the $2^1$ states ($ <\uparrow, \downarrow|, <\downarrow, \uparrow|$) the system occupies.  The maximal ignorance possible is 2 bits if the $2^2$ states ($<\downarrow, \uparrow|$, $<\downarrow, \uparrow|$, $<\uparrow, \uparrow|$, $<\downarrow, \downarrow|$) were possible. The mutual information of the two photons' polarizations is the same as $S^i_{EE}$, i.e.
\begin{align*}
I = S^i_{EE,\gamma} + S^i_{EE,x} - S^i_{EE} = S^i_{EE}.
\end{align*}
The maximum mutual information of $\gamma$ and $x$ for the mixed state, eqn. (\ref{imixed}), is 1 bit.  However, for a pure state, the maximum is I=2 bits.  A pure state gives more mutual information between the polarizations than a mixed state.  
If $I$ were zero, then the polarization of one photon does not convey information about the other photon.  e.g., the initial mutual information between $e^-$ and $x$ is zero, but the final mutual information after the scattering is non-zero (see section \ref{gamma}). This means the electron and witness particle become entangled after the scattering in spite of no direct interaction.\\

\section{Final reduced density matrices of photonic polarizations}\label{gamma and x}

\begin{figure}[htb]
\begin{center}
\includegraphics{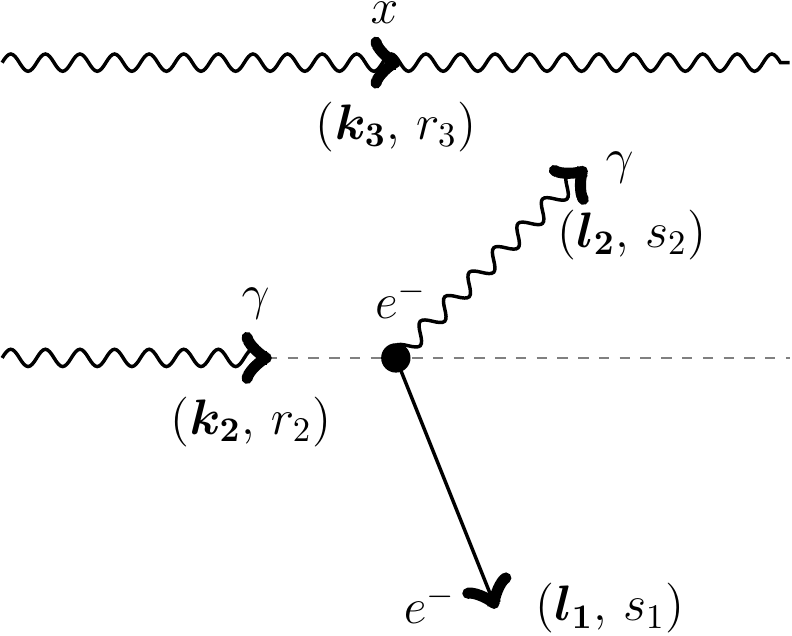}
\caption{$\gamma$ scatters off an $e^-$ at rest ($\vb*k_1=0$). $x$ is a witness particle that does not participate in the interaction.  ($\vb*{k_i}, r_i$) and ($\vb*{l_i},s_i$) are the initial and final momentum-polarization pairs, respectively.}
\label{compton1}
    \end{center}
\end{figure}

While keeping unitarity, a general derivation is provided for obtaining the final reduced density matrix of photonic polarizations entirely in terms of Feynman amplitudes. We assume an initially pure state.  If we assume an initally mixed state (see eqn. (\ref{imixed})), all $\cos\eta\sin\eta\exp^{\pm i\beta}$ terms in $\rho^f$ below would be dropped.  Figure \ref{compton1} shows the Compton scattering between $e^-$ and $\gamma$ with witness $x$ not participating in the interaction.  This is a two-particle scattering process, $AB\rightarrow AB$, while a third particle $C$ is the witness particle. The initial and final Hilbert spaces are divided as $\mathcal{H}_A \otimes\mathcal{H}_B\otimes\mathcal{H}_C$ since the states are spanned by a basis of a free Hamiltonian.   The three-particle state is written as
$|\vb*{k_1},r_1;\vb*{k_2},r_2;\vb*{k_3},r_3\rangle = |\vb*{k_1},r_1\rangle\otimes |\vb*{k_2},r_2\rangle\otimes|\vb*{k_3},r_3 \rangle,
$ where ($\vb*{k_i},r_i$) are the momentum-polarization pairs for the three particles.  The inner product of a state is
$\langle \vb*{p},r|\vb*{q},s\rangle$=
$2E_{\vb*{p}} (2\pi)^3\delta^{(3)}(\vb*p-\vb*q)\delta^{r,s}.$  The process for including other final states such as double-Compton scattering is mentioned below equation (\ref{reducedxgamma}).  

An initial state 
\bc
$|i\rangle = |e^-,\vb*{k_1},r_1\rangle$ (cos $\eta\ |\gamma,\vb*{k_2}, \uparrow \rangle |x,\vb*{k_3},\downarrow \rangle +e^{i\beta}$ sin $\eta\ |\gamma,\vb*{k_2},\downarrow\rangle |x,\vb*{k_3},\uparrow\rangle)$
\ec
has entanglement only between the two photons ($\gamma$ and $x$) via the entanglement parameter $\eta$. Particle labels in the kets are hereafter suppressed. If the electron is  unpolarized, the reduced density matrices in this section, $\rho^f_{\gamma,x}$ and $ \rho^f_{\gamma}$, should be preceded by $\frac{1}{2}\sum_{r_1}$, where $r_1$ is the electron's polarization.  In section \ref{Iex}, an unpolarized $e^-$ is used to calculate the final mutual information of the electron and the witness particle.  The case of a polarized electron is a little more challenging as explained in section \ref{Iex}.

Since the $S$-matrix is $S = 1+iT$, the final full density matrix for particles $e^-$, $\gamma$, and $x$ after the scattering is

\begin{align*}
\rho^f &= S|i\rangle \langle i|S^\dagger = |i\rangle \langle i| + (iT)|i\rangle \langle i| + |i\rangle \langle i|(-iT^\dagger) + (iT)|i\rangle \langle i|(-iT^\dagger)\\
\end{align*}

The normalization constant, $N$, of $\rho^i$ and $\rho^f$ are the same and equal to\me
\[N=\prod_{i=1}^3(2E_{\vb*{k_i}}V),\]\me
where $V=(2\pi)^3\ \delta^3(0)$ is the volume that remains to be regularized \cite{Peschanski}.  The normalized reduced density matrix for the photonic polarizations is had by tracing over all three particles' momenta and the $e^-$'s polarization, $r_1$, i.e.

\begin{align}\label{reducedxgamma}
\rho^f_{\gamma,x}\ &= \bf A + \bf B + \bf C,
\end{align}
where
\begin{align*}
\bf A &= \frac{1}{N}tr_{\vb*{k_1},\vb*{k_2},\vb*{k_3},r_1}(|i\rangle \langle i|),\\
\bf B &= \frac{1}{N}tr_{\vb*{k_1},\vb*{k_2},\vb*{k_3},r_1}(\ iT\ |i\rangle \langle i|-|i\rangle \langle i|\ iT^\dagger),\\
\bf C &= -\frac{1}{N}tr_{\vb*{k_1},\vb*{k_2},\vb*{k_3},r_1}(\ iT\ |i\rangle \langle i|\ iT^\dagger).\\
\end{align*}
Before tracing, the following completeness relation for two particle states is inserted both to the left of $iT$ and to the right of $iT^\dagger$ in $\bf B$ and $\bf C$.
\[I_{2-particles}= \sum_{s_1,\ s_2\ }\ \prod_{i=1}^2\int \frac{d^3\vb*{l_i}}{(2\pi)^32E_{\vb*{l_i}}}|\vb*{l_i},s_i\rangle\langle \vb*{\vb*{l_i}},s_i|\]
($\vb*{l_1},s_1$) and ($\vb*{l_2},s_2$) are the final momentum-polarization pairs for the scattered $e^-$ and $\gamma$, respectively.  For unitarity to be upheld, other processes should be included.  e.g. to include the three particle final state of double-Compton scattering ($e^-\gamma \rightarrow e^- \gamma \gamma$), let $I_{2-particles} \rightarrow I_{2-particles} + I_{3-particles}$.  The latter would leave \textbf{A} and \textbf{B} above unchanged after tracing.  However, \textbf{C} would change.  Its argument would consist of a sum of two particle states, three particle states, and a mixture.  But, after tracing over all the momenta and electronic polarization, what remains is a $matrix\ direct\ sum$ of photonic- polarization matrices deriving from the two particle final state and three particle final state. For ease of calculation, $I_{3-particles}$ is suppressed, but mentioned again at the end of the derivations.\\

 When performing the trace in $\bf B$, the term $\frac{\cos^2\eta}{(2E_{\vb*{k_1}} V) (2E_{\vb*{k_2}} V)}|\uparrow, \downarrow\rangle \langle \uparrow, \downarrow|$ has the coefficient
\begin{align*}
 \langle \vb*{k_1},r_1;\vb*{k_2},\uparrow | (iT) |\vb*{k_1},r_1; \vb*{k_2}\uparrow \rangle\ +\ h.c.\ \ &=\\
&\\
 (2\pi)^4\ \delta^4(0)\Big(iM(\vb*{k_1},r_1;\vb*{k_2},\uparrow\  \rightarrow\ \vb*{k_1},r_1;\vb*{k_2},\uparrow)\ +\ h.c.\Big)&=\\
 &\\
-(2\pi)^4\ \delta^4(0)\Big(\ 2\ Im(M(\vb*{k_1},r_1;\vb*{k_2},\uparrow\  \rightarrow\ \vb*{k_1},r_1;\vb*{k_2},\uparrow))\ \Big) &=\\
 &\\
-(VT)\ 2E_{\vb*{k_1}}\ 2E_{\vb*{k_2}}\ \sum_f \sigma(\vb*{k_1},r_1;\vb*{k_2},\uparrow\  \rightarrow\ f).
\end{align*}
In the last step, $VT=(2\pi)^4 \delta^4(0)$ while the optical theorem relates the imaginary part of the forward scattering amplitude to the total cross-section $\sum_f \sigma$ \cite{peskin}.  The forward amplitude contribution is important since it corresponds to a peak in the differential cross-section in the laboratory frame of Compton scattering.  It is assumed that the electron is initially at rest, otherwise a relative velocity factor, $|\upsilon_{e^-} - \upsilon_{\gamma}|$, would appear.\\  

After tracing,
\begin{flalign}\label{a}
 \ \bf A=&\indent \cos^2\eta|\uparrow,\downarrow \rangle \langle \uparrow,\downarrow| + \sin^2\eta|\downarrow,\uparrow \rangle \langle \downarrow,\uparrow|&&\\
&+ \Big(\cos \eta\ \sin \eta\ e^{i\beta}|\downarrow,\uparrow \rangle \langle \uparrow,\downarrow| +h.c.\Big).&&\nonumber
\end{flalign}

Defining the amplitude as
\bc
$M_{\vb*{k_1},r_1;\vb*{k_2},r_2}^{\vb*{l_1},s_1;\vb*{l_2},s_2}\ = M(\vb*{k_1},r_1;\vb*{k_2},r_2\  \rightarrow\ \vb*{l_1},s_1;\vb*{l_2},s_2),$
\ec
then $\bf B$ has the form

\begin{align}\label{b}
\bf B=&\ -\frac{ \sum_f \sigma(\vb*{k_1},r_1;\vb*{k_2},\uparrow\  \rightarrow\ f)}{V/T} \cos^2\eta|\uparrow, \downarrow\rangle \langle \uparrow, \downarrow| \\
&\ -\frac{\sum_f \sigma(\vb*{k_1},r_1;\vb*{k_2},\downarrow\  \rightarrow\ f)}{V/T} \sin^2\eta|\downarrow, \uparrow\rangle \langle \downarrow, \uparrow|\nonumber\\
&\ +\frac{1}{2E_{\vb*{k_1}} 2E_{\vb*{k_2}}V/T}\Big(iM_{\vb*{k_1},r_1;\vb*{k_2},\uparrow}^{\vb*{k_1},r_1;\vb*{k_2},\downarrow}\ |\downarrow, \downarrow\rangle \langle \uparrow, \downarrow| + h.c.\Big)\ \cos^2\eta\nonumber\\
&\ +\frac{1}{2E_{\vb*{k_1}} 2E_{\vb*{k_2}}V/T}\Big(iM_{\vb*{k_1},r_1;\vb*{k_2},\downarrow}^{\vb*{k_1},r_1;\vb*{k_2},\uparrow}\ |\uparrow, \uparrow\rangle \langle \downarrow, \uparrow| + h.c.\Big)\ \sin^2\eta\nonumber\\
&\ +\frac{1}{2E_{\vb*{k_1}} 2E_{\vb*{k_2}}V/T}\sum_{s_2}\Big(ie^{i\beta} M_{\vb*{k_1},r_1;\vb*{k_2},\downarrow}^{\vb*{k_1},r_1;\vb*{k_2},s_2}\ |s_2, \uparrow\rangle \langle \uparrow, \downarrow| + h.c.\Big)\cos \eta\ \sin \eta\nonumber\\
&\ +\frac{1}{2E_{\vb*{k_1}} 2E_{\vb*{k_2}}V/T}\sum_{s_2}\Big(ie^{-i\beta} M_{\vb*{k_1},r_1;\vb*{k_2},\uparrow}^{\vb*{k_1},r_1;\vb*{k_2},s_2}\ |s_2, \downarrow\rangle \langle \downarrow, \uparrow| + h.c.\Big)\cos \eta\ \sin \eta\nonumber.
\end{align}
Each coefficient in $\bf B$ has a ratio of areas in natural units.  Each Feynman amplitude in $\bf B$ has an almost forward scattering amplitude with the polarization of $\gamma$ possibly changing after the scattering.  The last two lines in $\bf B$ are zero at tree level for a flipped polarization of the scattered photon.  Hence, higher order corrections may not be ignored.\\

$\bf C$ has the form

\begin{align}\label{c}
\bf C=&\indent  \frac{\cos^2\eta}{2E_{\vb*k_1}\ 2E_{\vb*k_2}\ V/T} \sum_{s_1,\ s_2,\ t_2}\ \prod_{i=1}^2|s_2,\downarrow\rangle \langle t_2,\downarrow|*\\
&\indent \int \frac{d^3l_i}{(2\pi)^32E_{\vb*l_i}}\ M_{\vb*{k_1},r_1;\vb*{k_2},\uparrow}^{\vb*{l_1},s_1;\vb*{l_2},s_2}\ (M_{ \vb*{k_1},r_1;\vb*{k_2},\uparrow}^{\vb*{l_1},s_1;\vb*{l_2},t_2})^\dagger\ (2\pi)^4\ \delta^4(k_1+k_2-l_1-l_2)\nonumber\\
&\ + \frac{\sin^2\eta}{2E_{\vb*k_1}\ 2E_{\vb*k_2}\ V/T} \sum_{s_1,\ s_2,\ t_2}\ \prod_{i=1}^2\ |s_2,\uparrow\rangle \langle t_2,\uparrow|*\nonumber\\
&\indent \int \frac{d^3l_i}{(2\pi)^32E_{\vb*l_i}}\ M_{\vb*{k_1},r_1;\vb*{k_2},\downarrow}^{\vb*{l_1},s_1;\vb*{l_2},s_2}\ (M_{ \vb*{k_1},r_1;\vb*{k_2},\downarrow}^{\vb*{l_1},s_1;\vb*{l_2},t_2})^\dagger\  (2\pi)^4\ \delta^4(k_1+k_2-l_1-l_2)\nonumber\\
&\ + \Big(\frac{\cos\eta\ \sin\eta\ e^{i\beta}}{2E_{\vb*k_1}\ 2E_{\vb*k_2}\ V/T}\ \sum_{s_1,s_2,t_2}\ \prod_{i=1}^2\ |s_2,\uparrow\rangle \langle t_2,\downarrow| *\nonumber\\
&\indent \int \frac{d^3l_i}{(2\pi)^32E_{\vb*l_i}}\ M_{\vb*{k_1},r_1;\vb*{k_2},\downarrow}^{\vb*{l_1},s_1;\vb*{l_2},s_2}\ (M_{ \vb*{k_1},r_1;\vb*{k_2},\uparrow}^{\vb*{l_1},s_1;\vb*{l_2},t_2})^\dagger\  (2\pi)^4\ \delta^4(k_1+k_2-l_1-l_2)\nonumber\\
&\ +  h.c.\nonumber \Big).
\end{align}
The final reduced density matrix, $\rho^f_{\gamma,x}=\bf A+\bf B+\bf C$, has a trace of one and is hermetian after the interaction.  Since a factor \large $\frac{1}{|\upsilon_{e^-}\ -\ \upsilon_{\gamma}|}$  \normalsize is missing, $\rho^f_{\gamma,x}$ is not manifestly Lorentz invariant with respect to collinear boosts. After diagonalization of the matrix, the von Neumann or Shannon entropy can be calculated via $-tr(\rho\ \log\ \rho)$.\\

The coefficients of $\cos^2\eta\ |\uparrow,\downarrow\rangle \langle \uparrow,\downarrow|$ from $\bf A, \bf B,$ and $\bf C$ above add to 
\begin{align}\label{optical}
(1-\ \frac{\sum_f \sigma(\vb*{k_1},r_1;\vb*{k_2},\uparrow\  \rightarrow\ f)}{V/T} +\ \frac{\sum_{s_1} \sigma(\vb*{k_1},r_1;\vb*{k_2},\uparrow\  \rightarrow\ s_1; \uparrow)}{V/T} ) 
\end{align}
and looks like a probability for the scattered photon's polarization not to flip if $V/T$ is the available area for scattering.  In the COM frame for Compton scattering, $V/T$ in equation (\ref{optical}) becomes
\begin{align}\label{ansatz}
\frac{V/T}{(1+\beta)}.    
\end{align}
$\beta$ is the ratio of the $e^-$ momentum to its energy in the COM frame.  This means that at high energies, the ostensible area for scattering is cut in half.  In other words, at a very high energy, the photon can not $see$ or scatter from the other side of the electron.  At low energies or large wavelenths, the photon can $see$ the  $e^-$'s other side.  This is a quantum mechanical effect in scattering \cite{sakurai}.  For an unpolarized electron, append $\frac{1}{2}\sum_{r_1}$ to $\rho^f_{\gamma,x}$ if the electronic initial density matrix is $\rho^i_e = \frac{1}{2}\sum_{r_1}|r_1\rangle\langle r_1|$.  

Tracing $\rho^f_{\gamma,x}$ over $\gamma$'s polarization gives particle $x$'s reduced density matrix, which is equation (\ref{reducedx}).  Tracing $\rho^f_{\gamma,x}$ over $x$'s polarization, $\gamma$'s reduced density matrix is 
\begin{align}\label{rhofg}
\rho^f_{\gamma}\ &= \bf D + \bf E,
\end{align}
where

\begin{align*}
\bf D=&\indent (1-\frac{ \sum_f \sigma(\vb*{k_1},r_1;\vb*{k_2},\uparrow\  \rightarrow\ f)}{V/T})\cos^2\eta|\uparrow \rangle \langle \uparrow|\\
&\ + (1-\frac{\sum_f \sigma(\vb*{k_1},r_1;\vb*{k_2},\downarrow\  \rightarrow\ f)}{V/T})\sin^2\eta|\downarrow \rangle \langle \downarrow|\\
&\ +\frac{1}{2E_{\vb*{k_1}} 2E_{\vb*{k_2}}V/T}\Big(iM_{\vb*{k_1},r_1;\vb*{k_2},\uparrow}^{\vb*{k_1},r_1;\vb*{k_2},\downarrow}\ |\downarrow\rangle \langle \uparrow| + h.c.\Big)\ \cos^2\eta\\
&\ +\frac{1}{2E_{\vb*{k_1}} 2E_{\vb*{k_2}}V/T}\Big(iM_{\vb*{k_1},r_1;\vb*{k_2},\downarrow}^{\vb*{k_1},r_1;\vb*{k_2},\uparrow}\ |\uparrow\rangle \langle \downarrow| + h.c.\Big)\ \sin^2\eta,\\
&\\
\bf E=&\ \frac{\cos^2\eta}{2E_{\vb*k_1}\ 2E_{\vb*k_2}\ V/T} \sum_{s_1,\ s_2,\ t_2}\ \prod_{i=1}^2|s_2\rangle \langle t_2|*\\
&\indent \int \frac{d^3l_i}{(2\pi)^32E_{\vb*l_i}}\ M_{\vb*{k_1},r_1;\vb*{k_2},\uparrow}^{\vb*{l_1},s_1;\vb*{l_2},s_2}\ (M_{ \vb*{k_1},r_1;\vb*{k_2},\uparrow}^{\vb*{l_1},s_1;\vb*{l_2},t_2})^\dagger\ (2\pi)^4\ \delta^4(k_1+k_2-l_1-l_2)\\
&\ + \frac{\sin^2\eta}{2E_{\vb*k_1}\ 2E_{\vb*k_2}\ V/T} \sum_{s_1,\ s_2,\ t_2}\ \prod_{i=1}^2\ |s_2\rangle \langle t_2|*\\
&\indent \int \frac{d^3l_i}{(2\pi)^32E_{\vb*l_i}}\ M_{\vb*{k_1},r_1;\vb*{k_2},\downarrow}^{\vb*{l_1},s_1;\vb*{l_2},s_2}\ (M_{ \vb*{k_1},r_1;\vb*{k_2},\downarrow}^{\vb*{l_1},s_1;\vb*{l_2},t_2})^\dagger\  (2\pi)^4\ \delta^4(k_1+k_2-l_1-l_2).
\end{align*}
Notice $\rho^f_\gamma$ above is the same for an initially pure or mixed state since no $\cos\eta\sin\eta \exp^{\pm i\beta}$ terms appear in $\bf D$ and $\bf E$.  Keeping terms up to order $\alpha^2$ $\sim$ $e^4$ and having an arbitrary entanglement parameter $\eta$, $\rho^f_\gamma$ would require an expansion beyond tree level in the third and fourth line of $\bf{D}$ above.  Alternatively, only tree level terms are required if $\eta=\pi/4$ and the $general$ optical theorem is used.

e.g., using the optical theorem \cite{peskin} and letting $\eta=\pi/4$, the coefficient of $\frac{1}{2\ 2E_{\vb*k_1} 2E_{\vb*k_2} V/T}|\downarrow\rangle \langle \uparrow|$ in $\bf{D}$ becomes
\begin{align*}
&i\big(M_{\vb*{k_1},r_1;\vb*{k_2},\uparrow}^{\vb*{k_1},r_1;\vb*{k_2},\downarrow} - (M_{\vb*{k_1},r_1;\vb*{k_2},\downarrow}^{\vb*{k_1},r_1;\vb*{k_2},\uparrow})^\dagger\Big)=\\
&-\sum_{s_1, s_2} \prod_{i=1}^2
\ \int \frac{d^3l_i}{(2\pi)^32E_{\vb*l_i}}\ M_{\vb*{k_1},r_1;\vb*{k_2},\uparrow}^{\vb*{l_1},s_1;\vb*{l_2},s_2}\ (M_{ \vb*{k_1},r_1;\vb*{k_2},\downarrow}^{\vb*{l_1},s_1;\vb*{l_2},s_2})^\dagger\  (2\pi)^4\ \delta^4(k_1+k_2-l_1-l_2).
\end{align*}
A similar result occurs for $|\uparrow\rangle \langle \downarrow|$ in $\bf{D}$.  

Including three particle states such as double- Compton scattering or pair production in the density matrix would only affect \textbf{C} and \textbf{E} in equations (\ref{reducedxgamma}) and (\ref{rhofg}) because of the tracing.  e.g. this is accomplished by taking \textbf{C} and tacking on a momentum-polarization pair ($\vb*l_3,t_3$) to the superscripts of its Feynman amplitudes and an overall sum $\sum_{t_3}$. Then, add this result to the original \textbf{C} and do the same for \textbf{E}.  In this way, unitary or approximately unitary density matrices are calculated.

The presented procedure for calculating density matrix formulas preserves unitarity. In the subsequent section, unitarity approximately holds in that only the low energy Compton scattering of the electron and photon is considered in the presence of a witness particle.  According to Thirring's theorem \cite{Denner}, the higher order QED corrections to Compton scattering cancel at low energies. Including an order $O(\alpha^3)$ interaction such as the double-Compton scattering process $e^-\gamma\rightarrow e^- \gamma \gamma$ plus loop corrections in $e^-\gamma\rightarrow e^- \gamma$ is left for a future work.

\section{Quantum information metrics}\label{gamma}

\begin{figure}[htb]
\begin{center}
\includegraphics{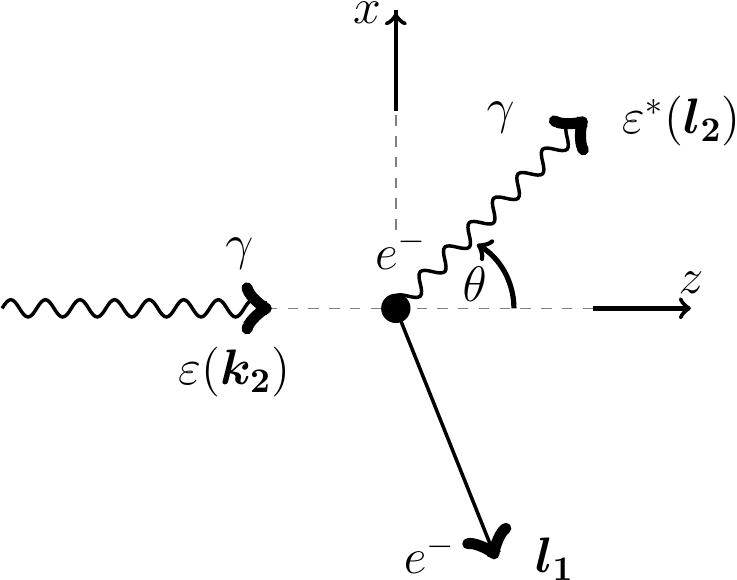}
\caption{$\gamma$ scatters off a polarized $e^-$ at rest. $\varepsilon(\vb*{k_2})$ and $\varepsilon^*(\vb*{l_2})$ are the initial and final polarization vectors of the photon. $\vb*{l_1}$ is the $e^-$'s final momentum.  Although entangled with $\gamma$, the witness particle $x$ is not depicted and does not participate in the interaction.}
\label{compton2}
    \end{center}
\end{figure}

The change in $\gamma$'s von Neumann entanglement entropy as well as the mutual information  between $e^-$ and $x$ are calculated in the low energy limit of Compton scattering.  If the $e^-$ is bound to an atom, assume the incident photon's energy is above that required for the photoelectric effect and well below that required for electron-positron pair production. The photon, $\gamma$, scatters from an $e^-$ at rest (see Figure \ref{compton2}).  The initial $e^-$ is polarized along $\hat{k_2}$ or the $z-$axis and has an inital momentum $k_1^\mu = (m,0,0,0)$. $\gamma$ has the following initial and final momenta and polarizations.  

\begin{align}\label{ke}
k_2^\mu &= (k_2,0,0,k_2), \hspace{60pt} l_2^\mu = (l_2,l_2 \sin\theta,0,l_2 \cos\theta),\\
\varepsilon^\mu(\vb*{k_2})&=\frac{1}{\sqrt{2}}(0,1,\pm i,0), \hspace{28pt} \varepsilon^{*\mu}(\vb*{l_2})=\frac{1}{\sqrt{2}}(0,\cos\theta,\mp i,-\sin\theta)\nonumber
\end{align}
The upper and lower signs in the above polarization vectors refer to right and left-handed polarizations, respectively.  An electron with momentum $\vb*p$ has the Dirac spinor
\begin{align*}
\large u(\vb*p)=
\begin{pmatrix}
\phi \\
  \frac{\vb*\sigma \cdot \vb*p}{E_{\vb*p}+m} \phi
\end{pmatrix},
\end{align*}
where the two-spinor $\phi$ is (1,0) or (0,1) and $\sigma_i$ is the Pauli matrix.

The Feynman amplitude for the process shown in Figure \ref{compton2} has an $s$ and $u$ channel \cite{peskin} and is given by

\begin{align}\label{su}
i \mathcal{M}= \indent &\bar{u}(\vb*l_1)
   (-ie \slashed{\epsilon}^* (\vb*l_2)
   )\ i\frac{(\slashed{k}_1 -\slashed{k}_2+m)}{2 k_1 \cdot k_2}\
   (-ie \slashed{\epsilon} (\vb*k_2))
  u(\vb*k_1)\\
  \nonumber\\
+\ &\bar{u}(\vb*l_1)
   (-ie \slashed{\epsilon} (\vb*k_2)
   )\ i\frac{(\slashed{k}_1 -\slashed{l}_2+m)}{-2 k_1 \cdot l_2}\
   (-ie \slashed{\epsilon}^* (\vb*l_2))
  u(\vb*k_1).\nonumber
\end{align}

\subsection{ Scattered photon's entropy}\label{5.1}

We assume an initial maximally entangled state ($\eta=\pi/4)$ between $\gamma$ and $x$, i.e.
\begin{align*}
|i\rangle = |e^-,r_1\rangle\  (\frac{1}{\sqrt{2}}|\gamma, \uparrow \rangle |x,\downarrow \rangle +e^{i\beta}  \frac{1}{\sqrt{2}}\ |\gamma,\downarrow\rangle |x,\uparrow\rangle).
\end{align*}
$\gamma$ has the initial reduced density matrix 
 \begin{align}\label{rhoig}
 \rho^i_\gamma = \frac{1}{2}(|\uparrow\rangle \langle \uparrow| + |\downarrow\rangle \langle \downarrow|)  
 \end{align}
 which is an initial unpolarized state.
From equation (\ref{rhoig}), the inital expected helicity of $\gamma$ along the $z$-axis is zero as are the Stokes parameters \cite{Landau}.  

 Using equations (\ref{rhofg} - \ref{su}) for an initial $e^-$ polarized along the $+z$-axis, the final reduced density matrix, $\rho_\gamma^f$, is 
\begin{equation}\label{rhoffg}
\rho_\gamma^f = 
\begin{pmatrix}
\frac{1}{2}(1 - \frac{\sigma_{T}}{V/T}\frac{3\omega}{2m}) & 0 \\
0 & \frac{1}{2}(1 + \frac{\sigma_{T}}{V/T}\frac{3\omega}{2m})  
\end{pmatrix},
\end{equation}
where $\sigma_T=\frac{8\pi \alpha^2}{3m^2}$ is the (unpolarized) Thomson scattering cross section \cite{peskin}.  The eigenvalues of $\rho_\gamma^f$ are $\frac{1}{2}(1 \mp \frac{\sigma_{T}}{V/T}\frac{3\omega}{2m})$ and correspond to matrix elements $|\uparrow\rangle \langle \uparrow|$ and $|\downarrow \rangle \langle \downarrow|$, respectively.    
For an $e^-$ initially polarized in the opposite direction, the plus and minus signs in equation (\ref{rhoffg}) are reversed.  If the initial electron is unpolarized \footnote{Let an initial unpolarized electron have the density matrix $\rho^i_e = \frac{1}{2}\sum_{r_1}|r_1\rangle\langle r_1|$.} or if $\rho^f_\gamma$ had been only expanded up to $\alpha \sim e^2$ instead of $\alpha^2$, the expected helicity given below or Stokes parameters would be zero.     

Since the Stokes parameters are not all zero in equation (\ref{rhoffg}), this final state of $\gamma$ is a partially polarized state.  The scattered $\gamma$'s final expected helicity is the Stokes parameter
\begin{align}\label{helicity}
\langle\sigma_z \rangle = tr(\sigma_z\ \rho^f_\gamma) = -\frac{\sigma_{T}}{V/T}\frac{ 3\omega}{2m}=-\frac{4\pi r^2_e}{V/T}\frac{\lambda_e}{\lambda_\gamma},
\end{align}
where $r_e$ is the electron's classical radius.  The $\lambda$'s are the wavelengths of the initial electron and photon.  Therefore, the reduced density matrix and helicity can be written entirely in terms of lengths and areas. For an oppositely polarized $e^-$, the expected helicity is positive.   Since the Stokes parameter or helicity is less than or equal to one in magnitude, $V/T \geq \frac{3\omega \sigma_T}{2m}$.  In the center of mass frame, wherein both colliding particles have momentum $\omega$, the Stokes parameter becomes -$\frac{\sigma_T}{2V/T} \frac{\omega^3}{m^3}$, again assuming low energy scattering.  Since the scattered photon's Stokes' parameter is Lorentz invariant, $(V/T)_{Lab}>(V/T)_{CoM}$, higher energy scattering appears to reduce the accessible area.  

Unlike the witness particle's entropy (see section (\ref{witness})), the von Neumann entanglement entropy of $\gamma$'s polarization changes due to the Compton scattering.  The change is  
\begin{align}\label{seeg}
\Delta S_{EE,\gamma}=&\ S_{EE,\gamma}^f - S_{EE,\gamma}^i\\
=&-tr (\rho_\gamma^f\log \rho_\gamma^f) + tr (\rho_\gamma^i\ \log \rho_\gamma^i)\nonumber\\
=&- \frac{1}{2}(1-\frac{\sigma_{T}}{V/T}\frac{ 3\omega}{2m})\log(1-\frac{\sigma_{T}}{V/T}\frac{ 3\omega}{2m})\nonumber\\
& -\frac{1}{2}(1+\frac{\sigma_{T}}{V/T}\frac{ 3\omega}{2m})\log(1+\frac{\sigma_{T}}{V/T}\frac{ 3\omega}{2m}).\nonumber
\end{align}
Since $\gamma$ transitions from an unpolarized  to a partially polarized state, $\Delta S_{EE,\gamma}\leq 0$ means $\gamma$'s polarization becomes less uncertain.  It has not been shown that a subsystem's entropy in general is always non-decreasing \cite{Cheung}. 
In the limit $\omega \rightarrow 0$ or no scattering, $\Delta S_{EE,\gamma}$=0 as expected. Note that $S_{EE,\gamma}$ is not Lorentz invariant unless it includes both the polarization and momentum of $\gamma$ \cite{Fan2}.

\subsection{Mutual information between the electron and witness particle}\label{Iex}

Earlier in section (\ref{witness}), it was seen that the witness particle, $x$, undergoes no change in its entanglement entropy due to the the Compton scattering between $\gamma$ and the electron. Also, $x$ was not correlated with the electron before the scattering.  However, the electron and the witness particle become correlated after the scattering. This can be seen by calculating the final mutual information between the unpolarized electron and $x$. This requires the von Neumann entanglement entropies, $S^f_{EE,ex}$ and $S^f_{EE,x}$, by way of the reduced density matrices, $\rho^f_{ex}$ and $\rho^f_e$, as given in equations
(\ref{reducedex})-(\ref{rhofe}) in appendix \textbf{A}. The case of a polarized $e^-$ is more challenging since the Feynman amplitude must be expanded to include higher order perturbations.  In other words, the optical theorem can be exploited in the polarized case.

Terms in $\rho^f_{e,x}$ are kept up to order $\alpha^2 \sim e^4$ since the process is a low energy Compton scattering.  This gives a  density matrix of 
\begin{equation}\label{rhofex}
\rho_{e,x}^f = 
\begin{pmatrix}
a_{-} \cos^2\eta & 0 &0 & b_{-} \cos^2\eta \\
0 & a_{-} \sin^2\eta  & b_{+}\sin^2\eta\  & 0\\
\\
0 &  b_{+}\sin^2\eta & a_{+} \sin^2\eta & 0\\
\\
b_{-} \cos^2\eta  & 0 & 0 & a_{+} \cos^2\eta 
\end{pmatrix} ,
\end{equation}
where $a_{\pm}=\frac{1}{2}(1 \pm \frac{\sigma_{T}}{V/T}\frac{\omega^2}{2m^2})$
 and $b_\pm=\pm \frac{\sigma_{T}}{V/T}\frac{9\pi \omega^2}{128m^2}$, and $\sigma_T=\frac{8\pi r_e^2}{3}$ is the Thomson scattering cross section for a free electron that scatters classical electromagnetic radiation.  $\omega$ and $m$ are the initial energy of the photon and rest mass of the electron, respectively. 
 
 The matrix elements of equation (\ref{rhofex}) from top left to bottom right along the diagonal are $|\downarrow\downarrow\rangle\langle \downarrow\downarrow|,\ |\uparrow \uparrow\rangle\langle \uparrow \uparrow|,\ |\downarrow \uparrow\rangle\langle \downarrow \uparrow|,\ |\uparrow \downarrow\rangle\langle \uparrow \downarrow|$.  Notice the matrix is Hermetian and the trace is one. Its four real eigenvalues, $\lambda_i$, are
\begin{align*}
\lambda_{1,2}=\frac {\cos^2\eta}{2}\  (1 \pm .67\frac {\sigma_T}{V/
T}\frac{\omega^2} {m^2} ),\indent \lambda_{3,4}=\frac {\sin^2\eta}{2}\  (1 \pm .67\frac{\sigma_T}{V/T}\frac{\omega^2}{m^2}).    
\end{align*}
and the final von Neumann entanglement entropy is $S^f_{EE,ex} = -\sum_{\lambda_i}tr\lambda_i \log \lambda_i$ while the initial entropy is $S^i_{EE,ex} = -\cos^2\eta \log \cos^2\eta -\sin^2\eta \log \sin^2\eta$.  When both $\omega \rightarrow 0$ and $\eta= \pi/4$ or $3\pi/4$, $\rho^f_{ex}$'s eigenvalues each equal $1/4$.  This would give an entanglement entropy of $S^f_{EE,ex} = 2$ bits, corresponding to four equally likely states for the electron and witness particle: $|\downarrow \downarrow\rangle,\ |\uparrow \uparrow\rangle,\  |\downarrow \uparrow\rangle,\ |\uparrow \downarrow\rangle$.

When calculating the change in entanglement entropy,  
\begin{align}\label{dsex}
\Delta S_{EE,ex}=&- \frac{1}{2}(1 + .67\frac{\sigma_T}{V/T}\frac{\omega^2}{m^2})\log (1 + .67\frac{\sigma_T}{V/T}\frac{\omega^2}{m^2})\\
& - \frac{1}{2}(1 - .67\frac{\sigma_T}{V/T}\frac{\omega^2}{m^2})\log (1 - .67\frac{\sigma_T}{V/T}\frac{\omega^2}{m^2}),\nonumber 
\end{align}    
the entanglement parameter, $\eta$, disappears.  

After tracing the latter density matrix over $x$'s polarizations, the electronic polarization's final reduced density matrix and change in entropy are
\begin{align}\label{sfe}
\rho_{e}^f =& 
\begin{pmatrix}
\frac{1}{2}(1 -\cos2\eta\  \frac{\sigma_{T}}{V/T}\frac{\omega^2}{2m^2}) &  -\cos2\eta\ \frac{\sigma_{T}}{V/T}\frac{9\pi \omega^2}{128m^2}  \\
-\cos2\eta\ \frac{\sigma_{T}}{V/T}\frac{9\pi \omega^2}{128m^2} & \frac{1}{2}(1 +\cos2\eta\  \frac{\sigma_{T}}{V/T}\frac{\omega^2}{2m^2})
\end{pmatrix},\nonumber \\ & \nonumber\\
\Delta S_{EE,e} =&- c_{-}\log c_{-} -c_{+}\log c_{+} - 1,\nonumber\\
=&-\frac{1}{2}(1+.67\cos2\eta\ \frac{\sigma_T}{V/T}\frac{\omega^2}{m^2})\log(1+.67\cos2\eta\ \frac{\sigma_T}{V/T}\frac{\omega^2}{m^2})\nonumber\\
&-\frac{1}{2}(1-.67\cos2\eta\ \frac{\sigma_T}{V/T}\frac{\omega^2}{m^2})\log(1-.67\cos2\eta\ \frac{\sigma_T}{V/T}\frac{\omega^2}{m^2})
\end{align}
where $c_{\pm}=\frac{1}{2}(1\pm.67\cos2\eta\ \frac{\sigma_T}{V/T}\frac{\omega^2}{m^2})$ are the eigenvalues of $\rho^f_e$ and $S^i_{EE,e}=1$.

\begin{figure}[htb]
\begin{center}
\includegraphics[width=\textwidth]{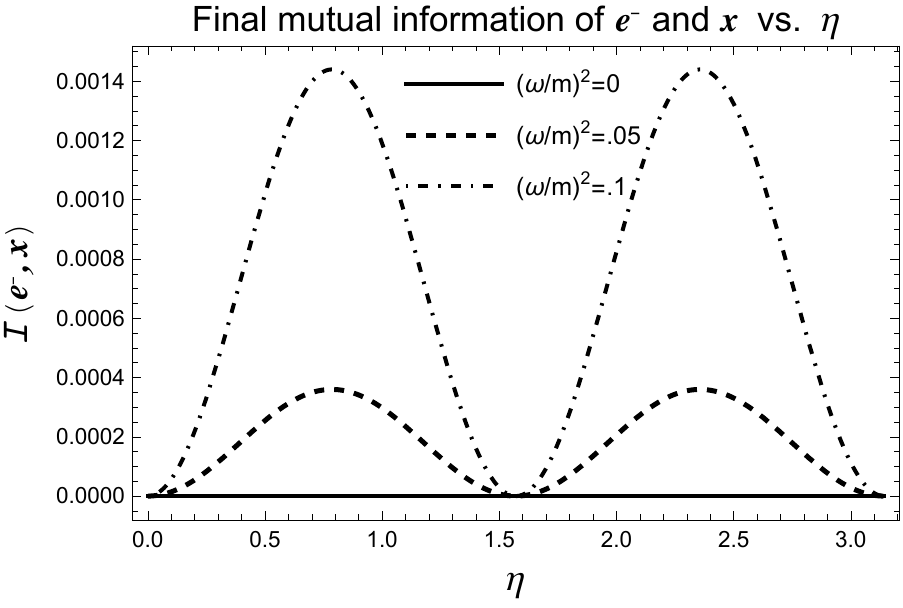}
\caption{Two photons ($\gamma$ and $x$) are assumed initially entangled via an entanglement parameter $\eta$.  Since the final mutual information $I(e^-,x)$ exceeds zero, the Compton scattering between an unpolarized $e^-$ and $\gamma$ serves as a proxy for the witness particle $x$ to interact with the electron.  $\omega/m$ is the ratio between the inital energies of $\gamma$ and the electron.}
\label{mutualgraph}
    \end{center}
\end{figure}

The only difference between $\Delta S_{EE,e}$ and $\Delta S_{EE,ex}$ above is a factor of $\cos2\eta$. Using these two equations (\ref{dsex}) and (\ref{sfe}), the mutual information $I(e^-,x)$ of $e^-$ and $x$'s polarizations is
\begin{align*}
\Delta I(e^-,x)=&I^f(e^-,x)-I^i(e^-,x)\\
=&\Delta S_{EE,e} + \Delta S_{EE,x} - \Delta S_{EE,ex}\\
\geq& 0.
\end{align*}
Since the initial mutual information $I^i(e^-,x)$ was zero, the Compton scattering causes $e^-$ and $x$ to become correlated in spite of not interacting directly.  The scattering of $\gamma$ from $e^-$ serves as a proxy for $x$ interacting with $e^-$ due to $\gamma$ and $x$'s initial entanglement.  If the initial entanglement between $\gamma$ and $x$ disappears ($\eta\rightarrow 0,\pi/2, \pi$), then $\Delta I(e^-,x)=I^f(e^-,x)\rightarrow 0$ as expected. The maximum mutual information occurs at $\eta=\pi/4$ or $3\pi/4$.

For the purpose of having a graph of $I$, assume $V/T = 4\pi r_e^2$ (see Figure \ref{mutualgraph}).  This is motivated by equations (\ref{optical}) and (\ref{ansatz}), wherein the ostensible available area, $V/T$, is divided by two for a highly energetic CoM collision. Classically, the photon should have access to scattering from an area $2\pi r_e^2$.  At low energies or large wavelengths, the photon can access the entire area $4\pi r_e^2.$  Also, the factor $V/T$ appears in the scattered photon's expected helicity (equation \ref{helicity}), which is a physical observable. Currently, no formal regularization exists \cite{Peschanski}.

\section{Discussion}
Deriving density matrix formulas via a general procedure that preserves unitarity under scattering is presented.  This is done by keeping the forward scattering amplitude.  A low energy Compton scattering process with a witness particle is considered.  Initially, the witness particle is entangled with a photon that later undergoes Compton scattering. Calculations of final density matrices, von Neumann entropies, and the mutual information between the electron and a witness particle are given. The mutual information of the unpolarized electron and witness particle after the scattering is not zero. This means that the two particles are entangled even though the witness particle did not interact directly with the electron.  The effect of a polarized electron on the mutual information is left for a subsequent work as it requires higher order corrections to the Feynman amplitude.

The reduced density matrix elements, Stokes parameters, entropies, and mutual information are found to consist of a common ratio of areas.  This ratio is the Thomson scattering cross section to the ostensible accessible scattering area $V/T$. $V/T$ has not been formally regularized in the literature \cite{Peschanski}.  Given that the scattered photon's expected helicity is finite and a function of $V/T$ (see section \ref{5.1}), a regularization procedure should exist.     

In future work, it would also be interesting to see how the mutual information of the electron and witness particle changes at higher energies.  Recentely, the Compton scattering cross section was calculated up to order $O(\alpha^3)$ \cite{Lee}.  $O(\alpha^3)$ corrections can be significant as they account for 7.6\% and 23\% of the Compton cross section at energy scales of $E_{CM}=1$ GeV and  $E_{CM}=1$ TeV, respectively.  Calculations in this paper can be extended by including $O(\alpha^3)$ processes.  This would include the double-Compton scattering process $e^-\gamma \rightarrow e^- \gamma \gamma$.  According to the Bloch-Nordsieck theorem, loop corrections in $e^-\gamma \rightarrow e^- \gamma$ are needed for infrared divergences in the latter two processes to cancel  \cite{Bloch}.  One might expect the correlation between the electron and witness particle to lessen when including photon creation and pair production.

\begin{appendix}\label{appA}
\section{Density matrices of the electron and witness particle}
Photons $\gamma$ and $x$ are initially entangled via the parameter $\eta$.  $x$ is the witness particle while $\gamma$ and $e^-$ undergo Compton scattering.
Assume the electron is unpolarized.  Suppressing the momenta, the initial density matrix is 
\begin{align*}
\rho^i =& \Big(\frac{1}{\sqrt{2}}\sum_{r_1}\ |e^-,r_1\rangle\langle e^-,r_1|\Big)\otimes\\
&\indent \Big(\cos^2 \eta|\gamma,\uparrow;x,\downarrow\rangle  \langle \gamma,\uparrow;x,\downarrow|+\sin^2 \eta|\gamma,\downarrow;x,\uparrow\rangle  \langle \gamma,\downarrow;x,\uparrow|\nonumber\\
&\indent + \cos\ \eta\ \sin\ \eta\exp(-i\beta)\ |\gamma,\uparrow;x,\downarrow\rangle  \langle \gamma,\downarrow;x,\uparrow|+h.c.\Big).  
\end{align*}

Following the section (\ref{gamma and x}) procedure, the reduced density matrix of the two qubit system, i.e. the polarizations for
an electron and a witness particle, is \begin{align}\label{reducedex}
\rho^f_{e,x}\ &= \bf F + \bf G + \bf H,
\end{align}
where

\begin{align}\label{fgh}
\bf F=&\frac{1}{2}\sum_{r_1}\ (\cos^2\eta\ |r_1,\downarrow \rangle \langle r_1,\downarrow| + \sin^2\eta\ |r_1,\uparrow \rangle \langle r_1,\uparrow|),\\
&\nonumber \\
\bf G=&\ \frac{1}{2\cdot 2E_{\vb*{k_1}} 2E_{\vb*{k_2}}V/T}\sum_{r_1,s_1}\Big(\\
&\indent (i\cos^2\eta\ M_{\vb*{k_1},r_1;\vb*{k_2},\uparrow}^{\vb*{k_1},s_1;\vb*{k_2},\uparrow}\ |s_1, \downarrow\rangle \langle r_1, \downarrow| + h.c.)+\ \nonumber\\
&\indent (i\sin^2\eta\ M_{\vb*{k_1},r_1;\vb*{k_2},\downarrow}^{\vb*{k_1},s_1;\vb*{k_2},\downarrow}\ |s_1, \uparrow\rangle \langle r_1, \uparrow| + h.c.)+\ \nonumber\\
&\indent (i\cos \eta\ \sin \eta\ e^{i\beta} M_{\vb*{k_1},r_1;\vb*{k_2},\downarrow}^{\vb*{k_1},s_1;\vb*{k_2},\uparrow}\ |s_1, \uparrow \rangle \langle r_1, \downarrow| + h.c.)+\nonumber\\
&\indent (i\cos \eta\ \sin \eta\ e^{-i\beta} M_{\vb*{k_1},r_1;\vb*{k_2},\uparrow}^{\vb*{k_1},s_1;\vb*{k_2},\downarrow}\ |s_1, \downarrow \rangle \langle r_1, \uparrow| + h.c.)\nonumber\\
&\indent \Big),\nonumber\\
&\nonumber \\
\bf H=&\ \frac{\cos^2\eta}{2\cdot 2E_{\vb*k_1}2E_{\vb*k_2} V/T} \sum_{r_1,s_1, t_1,t_2}\ \prod_{i=1}^2|s_1,\downarrow\rangle \langle t_1,\downarrow|*\\
&\indent \int \frac{d^3l_i}{(2\pi)^32E_{\vb*l_i}}\ M_{\vb*{k_1},r_1;\vb*{k_2},\uparrow}^{\vb*{l_1},s_1;\vb*{l_2},t_2}\ (M_{ \vb*{k_1},r_1;\vb*{k_2},\uparrow}^{\vb*{l_1},t_1;\vb*{l_2},t_2})^\dagger\ (2\pi)^4\ \delta^4(k_1+k_2-l_1-l_2)\nonumber\\
&\ + \frac{\sin^2\eta}{2\cdot 2E_{\vb*k_1}2E_{\vb*k_2}V/T} \sum_{r_1,s_1, t_1,t_2}\ \prod_{i=1}^2\ |s_1,\uparrow\rangle \langle t_1,\uparrow|*\nonumber\\
&\indent \int \frac{d^3l_i}{(2\pi)^32E_{\vb*l_i}}\ M_{\vb*{k_1},r_1;\vb*{k_2},\downarrow}^{\vb*{l_1},s_1;\vb*{l_2},t_2}\ (M_{ \vb*{k_1},r_1;\vb*{k_2},\downarrow}^{\vb*{l_1},t_1;\vb*{l_2},t_2})^\dagger\  (2\pi)^4\ \delta^4(k_1+k_2-l_1-l_2)\nonumber\\
&\ + \Big(\frac{\cos\eta\ \sin\eta\ e^{i\beta}}{2\cdot 2E_{\vb*k_1} 2E_{\vb*k_2}V/T}\ \sum_{r_1,s_1,t_1,t_2}\ \prod_{i=1}^2\ |s_1,\uparrow\rangle \langle t_1,\downarrow| *\nonumber\\
&\indent \int \frac{d^3l_i}{(2\pi)^32E_{\vb*l_i}}\ M_{\vb*{k_1},r_1;\vb*{k_2},\downarrow}^{\vb*{l_1},s_1;\vb*{l_2},t_2}\ (M_{ \vb*{k_1},r_1;\vb*{k_2},\uparrow}^{\vb*{l_1},t_1;\vb*{l_2},t_2})^\dagger\  (2\pi)^4\ \delta^4(k_1+k_2-l_1-l_2)\nonumber\\
&\ +  h.c.\nonumber \Big).
\end{align}

After tracing $\rho^f_{ex}$ over $x$'s polarization, the electronic reduced density matrix is

\begin{align}\label{rhofe}
\rho^f_e =& \ \frac{1}{2}\Big( \sum_{r_1}|r_1\rangle \langle r_1|\\
&+\frac{1}{ 2E_{\vb*{k_1}}2E_{\vb*{k_2}}V/T}\sum_{r_1,s_1}\Big((i\cos^2\eta\ M_{ \vb*{k_1},r_1;\vb*{k_2},\uparrow}^{\vb*{k_1},s_1;\vb*{k_2},\uparrow}|s_1\rangle \langle r_1|+ h.c.)\nonumber\\
&\hspace{4.2cm}+(i\sin^2\eta\ M_{ \vb*{k_1},r_1;\vb*{k_2},\downarrow}^{\vb*{k_1},s_1;\vb*{k_2},\downarrow}|s_1\rangle \langle r_1|+ h.c.)\Big)\nonumber\\
&\ +\frac{\cos^2\eta}{ 2E_{\vb*k_1} 2E_{\vb*k_2} V/T} \sum_{r_1,s_1, t_1,t_2}\ \prod_{i=1}^2|s_1\rangle \langle t_1|*\nonumber\\
&\indent \int \frac{d^3l_i}{(2\pi)^32E_{\vb*l_i}}\ M_{\vb*{k_1},r_1;\vb*{k_2},\uparrow}^{\vb*{l_1},s_1;\vb*{l_2},t_2}\ (M_{ \vb*{k_1},r_1;\vb*{k_2},\uparrow}^{\vb*{l_1},t_1;\vb*{l_2},t_2})^\dagger\ (2\pi)^4\ \delta^4(k_1+k_2-l_1-l_2)\nonumber\\
&\ + \frac{\sin^2\eta}{ 2E_{\vb*k_1} 2E_{\vb*k_2} V/T} \sum_{r_1,s_1, t_1,t_2}\ \prod_{i=1}^2\ |s_1\rangle \langle t_1|*\nonumber\\
&\indent \int \frac{d^3l_i}{(2\pi)^32E_{\vb*l_i}}\ M_{\vb*{k_1},r_1;\vb*{k_2},\downarrow}^{\vb*{l_1},s_1;\vb*{l_2},t_2}\ (M_{ \vb*{k_1},r_1;\vb*{k_2},\downarrow}^{\vb*{l_1},t_1;\vb*{l_2},t_2})^\dagger\  (2\pi)^4\ \delta^4(k_1+k_2-l_1-l_2)\nonumber\\
&\indent \Big).\nonumber
\end{align}
\end{appendix}

\setcounter{secnumdepth}{0}

\section{Acknowledgements}

SS thanks Professor Enrique Galvez for his hospitality and useful discussions. SS also thanks Ms. Leia Francis for assistance with document preparation. This work was partly funded by a Colgate University Research Council grant.

\end{document}